\documentclass[fleqn,usenatbib]{mnras}

\usepackage{newtxtext,newtxmath}

\usepackage[T1]{fontenc}

\DeclareRobustCommand{\VAN}[3]{#2}
\let\VANthebibliography\thebibliography
\def\thebibliography{\DeclareRobustCommand{\VAN}[3]{##3}\VANthebibliography}

\usepackage{cases}

\usepackage{graphicx}	
\usepackage{amsmath}	
\usepackage{float}
\usepackage{pdflscape}

\title[Investigating the porosity of Enceladus]{Investigating the porosity of Enceladus}

\author[I. Kisvárdai et al.]{
Imre Kisvárdai,$^{1,2,3}$\thanks{E-mail: kisvardaiimre02@gmail.com (KTS)}
Bernadett D. Pál,$^{2,3}$
Ákos Kereszturi$^{2,3,4}$
\\
$^{1}$Eötvös Loránd University, Egyetem tér 1-3. 1053 Budapest, Hungary\\
$^{2}$Konkoly Thege Miklós Astronomical Institute, Research Centre for Astronomy and Earth Sciences, Konkoly-Thege Miklós Út 15-17. 1121 Budapest, Hungary\\
$^{3}$CSFK, MTA Centre of Excellence, Budapest, Konkoly Thege Miklós út 15-17., H-1121, Hungary\\
$^{4}$European Astrobiology Institute, 1, quai Lezay-Marnésia - BP 90015 67080 Strasbourg Cedex - France\\
}

\date{Accepted XXX. Received YYY; in original form ZZZ}

\pubyear{2023}

\begin{document}
\label{firstpage}
\pagerange{\pageref{firstpage}--\pageref{lastpage}}
\maketitle

\begin{abstract}
The interior of Enceladus, a medium sized icy moon of Saturn hosts hydrothermal activity and exhibits tidal heating and related geyser-like activity. There are major disagreements in the existing literature on the porosity of the interior, due to the different theoretical assumptions on which porosity related calculations were based. We present an application of experimental equations - derived for Earth - for icy planetary objects and Enceladus in particular. We chose a set of boundary values for our initial parameters from measured porosity values of chondrite samples as references, and calculated the porosity related values of Enceladus using various approaches. We present a comprehensive investigation of the effects of using these different porosity calculation methods on icy moons. With our most realistic approach we also calculated the same values for Earth and Mars for comparison. Our result for Enceladus is a minimum porosity of about 5\% at the centre of the body. For the total pore volume we estimated $1.51*10^7 km^3$ for Enceladus, $2.11*10^8 km^3$ for Earth and $1.62*10^8 km^3$ for Mars. Using the same method, we estimated the total pore surface area. From this we derived that the pore surface under a given $1 km^2$ area of the surface on Enceladus is about $1.37*10^9 km^2$, while for Earth this value is only $5.07*10^7 km^2$.
\end{abstract}




\section{Introduction}
\label{sec:intro}

The icy moons of the Solar System represent an excellent opportunity to investigate the liquid water related chemical processes within small icy bodies, that might support chemical changes and in theory influence the habitability potential for primitive chemotrophic lifeforms. Enceladus, with a surface gravity of only $0.113 ms^-2$ has an excessively porous interior, allowing large amounts of fluid to freely flow through and interact with the rock \citep{jacobson2006gravity, porco2006cassini}. This geologically active moon of Saturn probably hosts a subglacial ocean implied by the data of the Cassini mission \citep{thomas2016enceladus} and shown by various evolution models \citep{Neumann_2019, choblet2017powering}. Enceladus has been the topic of intense scientific discussion in recent years, because of a number of yet unexplained geological phenomena \citep{waite2017cassini, 2022MNRAS.517.3485K,Neumann_2019, choblet2017powering}. For example during a flyby the Cassini spacecraft detected sodium-salt-rich ice grains emitted by a plume of Enceladus \citep{hsu2015ongoing}. This suggests that the grains have been in contact with rock and recent studies show that they are connected to ongoing hydrothermal activity on the internal silicate surfaces \citep{sekine2015high}.\\

Although many studies have investigated the surface morphology and internal heating mechanisms of Enceladus, no comprehensive research has been conducted yet that focuses on the internal porosity of the icy moon, while this is an essential part of the geophysical aspects, that influences many processes in the interior  \citep{malamud2013modeling, Neumann_2019, choblet2017powering}. Processes that are influenced by porosity are for example the differentiation of the interior, the fluid flow and certain chemical reactions. The parameters of the internal structure of Enceladus can be approached using numerical models that consider compaction by self-gravity, possible initial surface porosity values and pore sizes. Contrary to some already existing structural evolution models of Enceladus, we aim to calculate the porosity of the planetary body using equations that are experimentally derived from porosity measurements conducted on rocky material on Earth \citep{schmoker1982carbonate, schmoker1988sandstone}. This method of calculation can be applied for any rocky planetary body as shown by \citet{clifford1993model}, who applied this approach for the Martian crust.   After calculating the porosity values of Enceladus we calculate the same parameters of Earth and of Mars using the same method for comparison and to study the effect of gravity variation on the porosity profiles. Although the interior structure and related material properties differ for the analysed bodies, the first approach presented here already could provide useful results for future, more sophisticated models. The investigation of the porosity of the interior of Enceladus can help us to better understand the permeability, the possibility of fluid related chemical processes, and related heat transport in the interior. The connections between the porosity and these properties is further discussed in Section \ref{sec:discussion} of this paper. These properties of the interior have a considerable effect on the habitability potential of Enceladus as they greatly influence the ambient temperature, redox potential and chemical composition of the whole interior \citep{trolard2022mineralogy}.\\

In this work we aim to estimate the possible porous volume and pore surface using various methods (such as an experimentally derived equation for the calculation of porosity and an interpolation method) describing the change of porosity with depth, and compare it to those of Earth and of Mars. We propose a model for the calculation of these parameters and describe it in Section \ref{sec:methods}. We present our results and the related figures in Section \ref{sec:results}. In Section \ref{sec:discussion} we discuss the potential geochemical and astrobiological relevance of the results. We also compare our results to other published structural models of Enceladus and discuss the possible cause of the agreements and disagreements. Lastly, in Section \ref{sec:conclusion} we summarize our results, evaluate their significance and discuss the possible implications and improvements of our work.

\section{Methods}
\label{sec:methods}

In this section we summarize the method of calculating the values of porosity related parameters. Such parameters are the variation of porosity with depth, the total pore volume and pore surface found in certain sections of the interior, and within the whole body. The following simplifications are implicated in our calculations:

\begin{enumerate}
    \item We consider each planetary body as a perfect sphere.
    \item We assume each individual pore void as a perfect sphere with uniform radius. Each of these pores are separated in space for simplification.
    \item We consider every location at the surface on each investigated planetary body to be of uniform porosity. This porosity value is the initial surface porosity.
\end{enumerate}

The simplifications described above have a variety of consequences on the realism of our model. As it describes a static state of the celestial bodies in time, it needs further improvement to be able to be used to investigate time dependent processes, such as fluid flow and mineral transport within the interior, as well as heat conduction, tectonics and fault dynamics. Furthermore, the phenomena of tiger stripes and cryovolcanic eruptions (discussed by \citet{brown2006composition} and \citet{souvcek2016effect}) and all other surface processes are omitted from our model. Each pore is filled with water in our model following the findings of \citet{postberg2011salt} for which we provided the density in Table \ref{tab:parameters}. The same densities are applied for the case of Earth and Mars too, which should be a sufficiently accurate approximation, because the general composition of the oceans of Earth and Noachian Mars was probably similar to that of Enceladus \citep{rodriguez2015martian, elderfield1996mid}. In reality, a certain fraction of the pores could be filled with gases or solid ice particles too, but it should have a small effect on the realism of our calculations; however, we are working on the implementation of this issue in a more developed version of the model. The neglect of the interconnection of the pores enlarges the amount of pore surface area there is compared to the real situation, so the calculated pore surfaces should be considered an upper boundary of the real values. This particular assumption has no other consequences regarding our results, because we do not model the fluid flow or any transport/convection processes in the interior. We also do not take the plastic deformation of the rock into consideration, because it would further complicate and add uncertainty to our calculations. However it should be incorporated in the case of larger bodies in the future using a more advanced version of this model that is still in development. For a realistic estimation of plastic deformation internal temperature values should be also considered. but it is necessary to compromise between the practicality of simple and effective models with simplified geometries \citep{2007Icar..188..345S, 2008Icar..197..211P} and the reality of costly numerical models \citep{choblet2017powering, Neumann_2019}. With plastic deformation due to the applied pressure, the density of the rock increases, but this effect most likely stays negligible, because even at the center of Enceladus, the pressure values are not high enough to compress the rock on a significant scale even if the plastic deformation were to be included.

\subsection{Porosity calculation with an empirical formula}
\label{sec:porosity}

The porosity (or void fraction) of a material is given by its void volume to total volume ratio, as the following equation shows \citep{trolard2022mineralogy}:

\begin{equation}
    \phi=\frac{V_{void}}{V_{total}}
	\label{eq:porosity}
\end{equation}

$\phi$ represents the porosity, $V_{void}$ the void volume and $V_{total}$ the total volume of the rock. The value of porosity of a rock is influenced by the amount of pressure applied to the rock: with increasing pressure the value of porosity decreases. This process is the compaction. We call a given rock compacted, if the porosity of the rock is 0. Geophysical studies show that the change of the porosity of rocks on Earth is best described by an exponential decline with depth, because of the increasing lithostatic pressure applied by the above layers of rock \citep{michalski2013groundwater, michalski2018martian}. This relationship can be applied to any rocky planetary body, as shown by \citet{clifford1993model}, not just to Earth. According to this model, the porosity at depth z is given by the following empirical equation \citep{athy1930density}:

\begin{equation}
    \phi(z)=\phi_{0}exp\left(\frac{-z}{k}\right)
	\label{eq:pordepth}
\end{equation}

where $\phi_0$ is the porosity at the surface and \textit{k} is the porosity decay coefficient. For the possible values of surface porosity we chose an interval constrained by the minima and maxima of the carbonaceous chondrite sample investigated by \citep{macke2011density}. The value of \textit{k} coefficient is also dependent on gravity, the density and the compressibility of the rock, as the following equation shows \citep{revil199910}:

\begin{equation}
    k=\frac{1}{\phi_0g\beta(\rho_b-\rho_f)}
	\label{eq:k}
\end{equation}

where $\beta$ represents the long term compressibility factor, $\rho_b$ is the bulk density of the rock, $\rho_f$ is the density of the fluid contained within the rock and g is the value of the gravitational acceleration at the surface of the body. In our calculations we only take the gravity dependency of k into consideration, because we want a general overview of the effect of gravity on the porosity of planetary bodies. The values of $\beta$, $\rho_b$ and $\rho_f$ are therefore considered to be constant, and the values of \textit{k} for Enceladus, for Earth and for Mars can be found in table \ref{tab:parameters}.\\

The interior of Enceladus is most likely at least partially differentiated. The differentiation probably results in three layers: the ice crust, the water ocean and the rocky core. In our model each layer has different density values, but within any given layer the density remains constant because of simplicity. We denote the densities of the layers as follows: $\rho_1$ is used for the rocky core, $\rho_2$ for the ocean and $\rho_3$ for the ice crust. \\

 First, we describe the method for the calculation of the value of the gravitational acceleration within a differentiated planetary body below. The value of the gravitational acceleration at a given depth has to be calculated as accurately as possible, because the porosity of a given planetary body is a function of this parameter. In these equations $M_1(r)$ denotes the mass of the spherical section of the inner core with radius \textit{r} around the centre, $M_2(r)$ denotes the mass of the spherical shell of water and $M_3$ denotes the mass of the spherical shell of ice. The variable \textit{r} denotes the current distance from the centre of the planetary body and $r_1$ and $r_2$ denotes the distance of the surface of the rocky core and the water layer from the centre. V(r) is the volume of a sphere with radius \textit{r}.\\

$
M_1(r) =
\begin{cases}

\rho_1*V(r) & \text{if }r<r_1\\
\rho_1*V(r_1) & \text{if }r_1<r

\end{cases}
$

$
M_2(r) =
\begin{cases}

0 & \text{if }r<r_1\\
\rho_2*(V(r)-V(r_1)) & \text{if }r_1<r<r_2\\
\rho_2*V((r_2)-V(r_1)) & \text{if }r_2<r

\end{cases}
$

$
M_3(r) =
\begin{cases}

0 & \text{if }r<r_2\\
\rho_3*(V(r)-V(r_2)) & \text{if }r_2<r

\end{cases}
$

\begin{equation}
\label{eq:M}
    M(r)=M_1+M_2+M_3
\end{equation}

After the calculation of \textit{M}, we can calculate the value of gravitational acceleration with the following equation:

\begin{equation}
\label{eq:g}
    g(r)=\frac{G*M}{r^2}
\end{equation}

The described model provides a non-linearly changing gravitational acceleration throughout the interior. By taking the value of this parameter at the surface of the rocky core, we can get a more accurate value for \textit{g}, which can be used in Equation \ref{eq:k}. With this more realistic value of \textit{k} we can calculate the porosity value at any given depth inside a planetary object with Equation \ref{eq:pordepth}.

 \subsection{Porosity calculation using porosity-pressure relations from measurement data}
 
 Approaching the task of modelling the porosity differently, here we took the pressure dependence into consideration instead of the empirical described subsection. The pressure-porosity relationship within the interior of a planetary body is rather complicated and we do not yet have accurate models describing this relation unfortunately. Therefore we interpolated the porosity-pressure data measured on Earth \citep{flemings2021concise} using the estimated core pressure of Enceladus (40 MPa) taken from the study of \citeauthor{choblet2017powering} and the porosity measured from rocks that were put under the same pressure. The other endpoint being the surface porosity $\phi_0$ of 40\% we assumed a direct relation between the rate of pressure change and the rate of porosity change and linearly connected the surface values to the core values. We omitted the effect of the hypothesized global ocean on top of the rocky core on the pressure conditions, because even a mass of water of the maximum height constrained by various authors \citep{souvcek2023radar, choblet2017powering, Neumann_2019} would not have a significant (more than 1\% difference) effect on the surface porosity of the rock. Data used for this method can be found in Table \ref{tab:inter}
 
 \begin{table}
     \centering
     \caption{Values taken from the studies of \citet{choblet2017powering, flemings2021concise} to interpolate the porosity and pressure data measured on silicate rocks inside Earth. The only value we derived in this table is the porosity of Enceladus at the given depth of 190000 m.}
     \begin{tabular}{lcccc}
          \hline
        & Surface porosity & Depth & Pressure & Porosity \\
           \hline
        Earth & 40\% & 2500 m & 40 MPa & 16\% \\
         Enceladus & 40\% & 190000 m & 40 MPa & 16\% \\
           \hline
     \end{tabular}
     \label{tab:inter}
 \end{table}

As a result of the pressure-porosity data interpolation, we got a porosity of 16\% at the core of Enceladus (at 190000 m depth from the surface of the rocky core).
Following the calculation of the porosity of the bodies at every depth, we can calculate their pore volumes.

\subsection{Total pore volume}
\label{sec:totalvolume}

The total volume occupied by pores inside a body is the total pore volume. This parameter also determines the amount of fluid or gas that can reside within the pore spaces of the rock. For the calculation of this parameter we derived the following equation by integrating the product of the surface area function of a sphere with radius r and Equation \ref{eq:pordepth} from the centre to the surface:

\begin{equation}
    V_\Sigma=\int_r^0\phi(z)4\pi (r-z)^2dz=4k\pi\phi_0(2k^2-2kz+z^2)exp\left(\frac{z-r}{k}\right)
	\label{eq:totalvolume}
\end{equation}

where z is the distance from the surface and $V_\Sigma$ is the total pore volume of the spherical shell with radius r and thickness z. This parameter gives us the volume in which fluid and gas can reside within the rock and possibly interact with it on pore surfaces.
 
\subsection{Total pore surface}
\label{sec:surface}

The total pore surface is the area of the combined surfaces of all the pores within a body. To calculate this value, we assumed, that all pores are perfect spheres, and the interval of the radii of these spheres are chosen using the minimum and maximum values of pore radii within the meteorite sample measured by \citet{friedrich2008pore} as boundary values. The relevant pore sizes are given in Table \ref{tab:parameters}. We derived the equation for the calculation of the total pore surface area of bodies from Equation \ref{eq:totalvolume}:

\begin{equation}
     A_\Sigma=\frac{V_\Sigma}{V_p}A_p
	\label{eq:PoreA}
\end{equation}

where $A_\Sigma$ is the total pore surface area and $V_p$ and $A_p$ are the volume and surface respectively of the individual pores for which we provided the minimum and maximum values of their radii in Table \ref{tab:parameters}. By using the upper limit of the initial surface porosity and the lower limit of the pore radii, we get the upper limit of the total pore surface area.

\subsection{Local concentration of pore volume and pore surface}
\label{sec:local}

In order to compare the amount of pore surfaces inside Enceladus, Earth and Mars, we calculate the value of the pore surface below a $1km^2$ area on each body. Please note that in our approach pore voids are considered to be separated, however in reality interconnected, communicating pore voids might be present. However, this simplification is necessary in order to be able to properly calculate the pore surface area of the planetary bodies.

\begin{equation}
     A_r=A_\Sigma\frac{1km^2}{4\pi r^2}
	\label{eq:1kmA}
\end{equation}

In this equation $A_r$ denotes the pore surface area under a $1km^2$ section of the surface of a planetary body and r is the radius of the body.

\subsection{Material properties and parameters}
\label{sec:model}

Material properties of Enceladus, of Earth, of Mars and of the rock are given in the following table:

\begin{table}
	\centering
	\caption{This table summarizes the material properties of Enceladus, of Earth and of Mars bodies and constraints on the properties of the rock, using measurements and research conducted by \citet{papike1976mare, thomas2016enceladus, travis2015keeping, rovira2022tides, vcadek2016enceladus, postberg2011salt}. Values of the porosity decay \textbf{coefficient} 'k' were derived using Equation \ref{eq:k}}
	\label{tab:parameters}
	\begin{tabular}{lccc} 
		\hline
	Parameter & Symbol & Value & Unit \\
	\hline
	Radius of Enceladus & $r_e$ & 252.1 & km  \\
	Radius of the core of Enceladus & $r_c$ & 190 & km  \\
    Surface gravity of the core of Enceladus & $g_e$ & 0.127 & $ms^{-2}$  \\
    Radius of Earth & $r_E$ & 6371 & km  \\
    Surface gravity on Earth & $g_E$ & 9.81 & $ms^{-2}$ \\
    Radius of Mars & $r_m$ & 3389.5 & km  \\
    Surface gravity on Mars & $g_m$ & 3.71 & $ms^{-2}$  \\
    Surface porosity & $\phi_0$ & 0.40 & -  \\
    Long term compressibility constant & $\beta$ & $1.41*10^{-7}$ & $Pa^{-1}$ \\
    Bulk density of the rock & $\rho_b$ & 2400 & $kgm^{-3}$  \\
    Density of the core of Enceladus & $\rho_1$ & 2400 & $kgm^{-3}$  \\
    Density of water & $\rho_2$ & 998 & $kgm^{-3}$ \\
    Density of ice & $\rho_3$ & 918 & $kgm^{-3}$  \\
    Maximum pore radius of the rock & $r_{max}$ & 500 & $\mu$m  \\
    Minimum pore radius of the rock & $r_{min}$ & 1  & $\mu$m \\
    Value of k on Enceladus & $k_e$ & $93186$ & m \\
    Value of k on Earth & $k_E$ & $1073$ & m \\
    Value of k on Mars & $k_m$ & $2820$ & m \\
		\hline
	\end{tabular}
\end{table}

\section{Results}
\label{sec:results}

\begin{figure}
\includegraphics[width=\columnwidth]{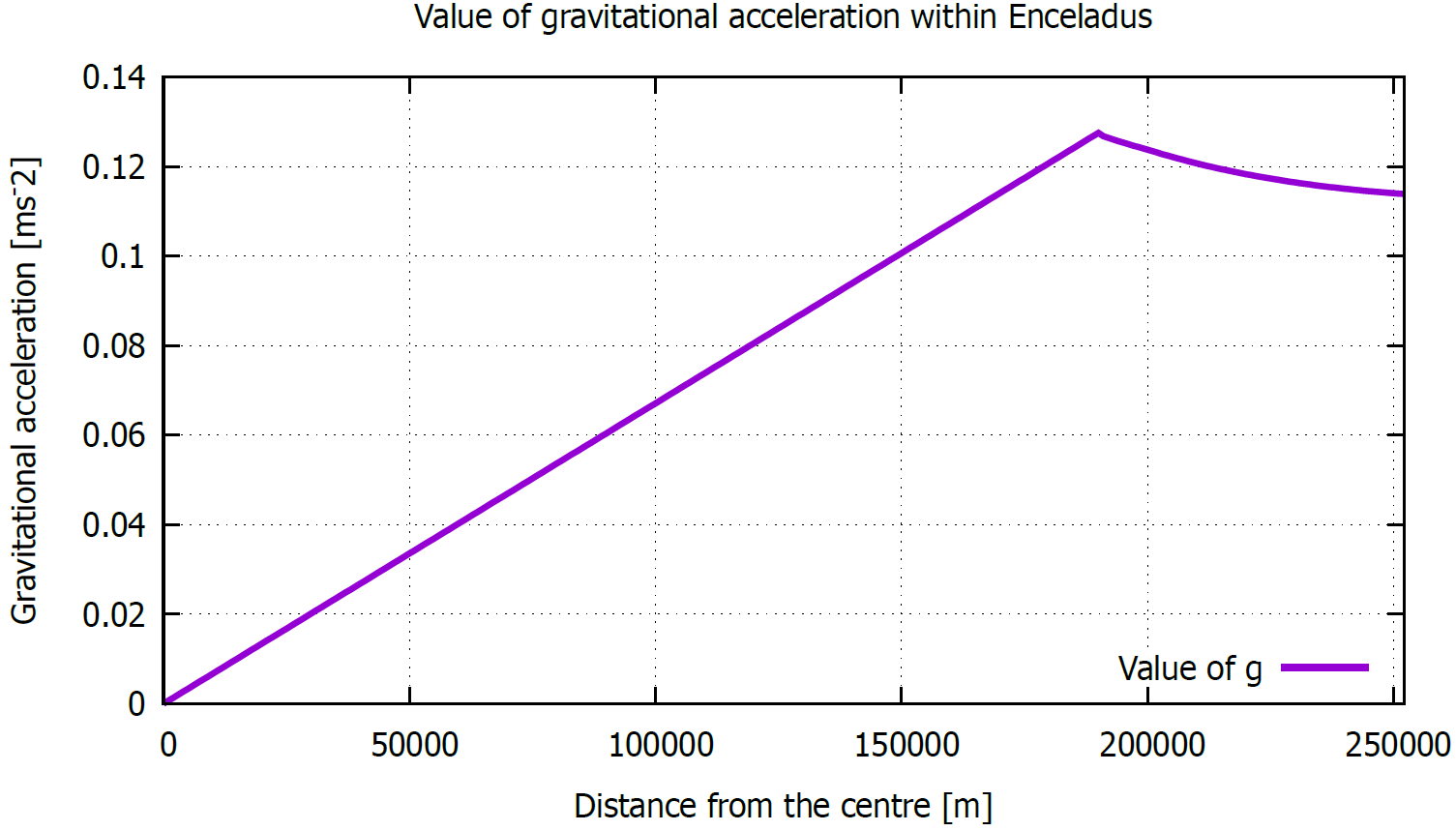}
    \caption{The value of the gravitational acceleration within the interior of Enceladus. It is evident, that the maximum of this value is at outer boundary of the rocky core, which is the densest region of the interior. For our porosity calculations in this paper, we use this maximum value in Equation \ref{eq:k}. This value is 0.127 $ms^{-2}$.}
    \label{fig:g}
\end{figure}

Calculating the value of the gravitational acceleration at every depth inside Enceladus with Equation \ref{eq:g} we got the result showcased on Figure\ref{fig:g}.

With this value, which is more accurate than the value of the gravitational acceleration measured at the surface of the icy crust, we calculated the value of k for Enceladus, which can be seen in Table \ref{tab:parameters} alongside the same parameters for Earth and Mars.\\

\begin{figure}
\includegraphics[width=\columnwidth]{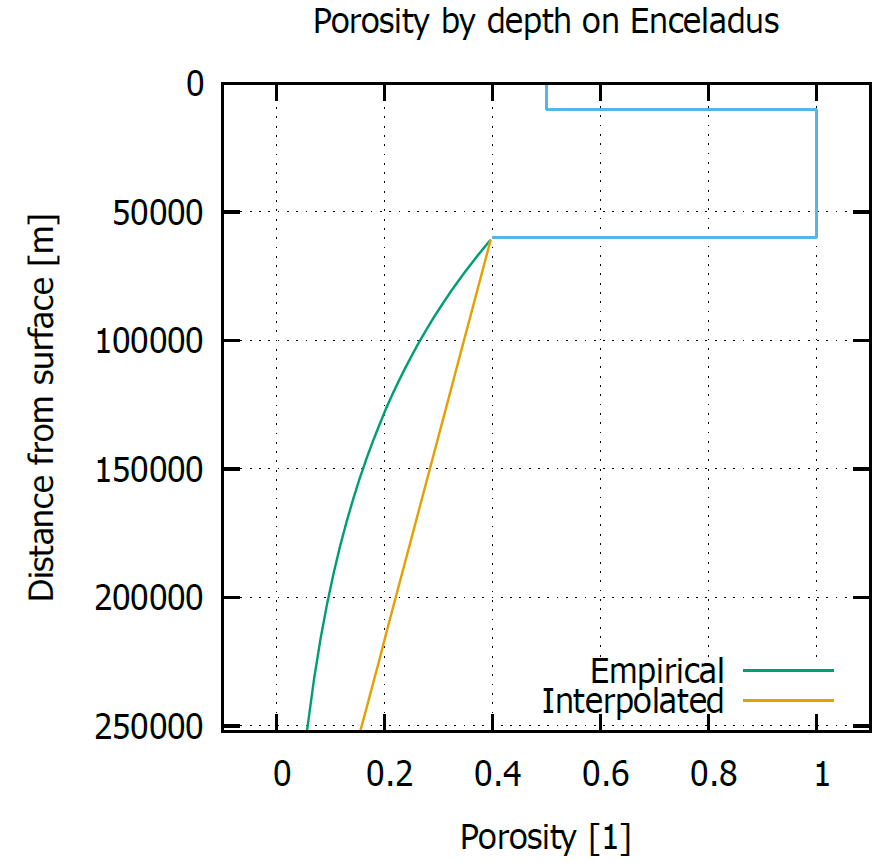}
    \caption{The blue line at the top right represents the porosity of the icy crust and the water layer, while the green and yellow lines represent the porosity calculated with the measurement based empirical and the interpolation method respectively. The porosity derived with the later method is substantially higher.}
    \label{fig:por}
\end{figure}

Using this k value for Enceladus we plotted the porosity curve of the planetary body derived with the empirical formula and the interpolation method too.
For comparison they are shown on the same plot on Figure \ref{fig:por}

\begin{figure}
	\centering
	    \includegraphics[width=\columnwidth]{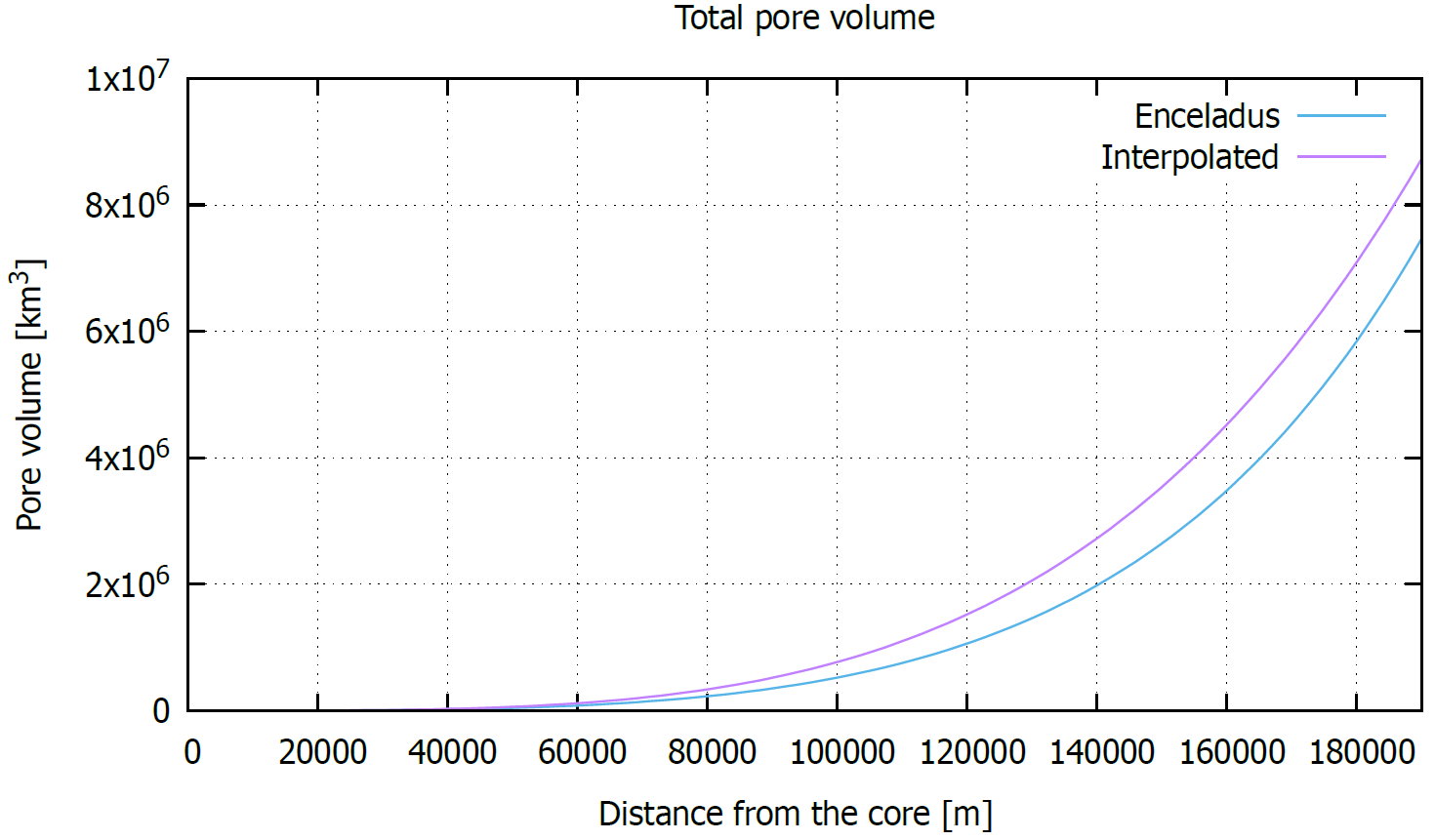}
	    \caption{This figure showcases the difference between the empirical and interpolation methods in the result of the total pore volume of Enceladus. While the empirical method results in a pore volume of $7.5*10^6 km^3$, the interpolation method gives us a volume of $8.8*10^6 km^3$}
	    \label{fig:V}
\end{figure}

The reason behind the difference in the porosity results between the two methods is that the empirical formula describes an exponential decline with depth, meanwhile the interpolation method describes a linear change. Having identical initial values, and roughly the same endpoints, this means that the interpolated results are going to be of higher value at each point of depth. The exact magnitude of this difference can be seen on Figure \ref{fig:V}\\

\begin{figure}
\includegraphics[width=\columnwidth]{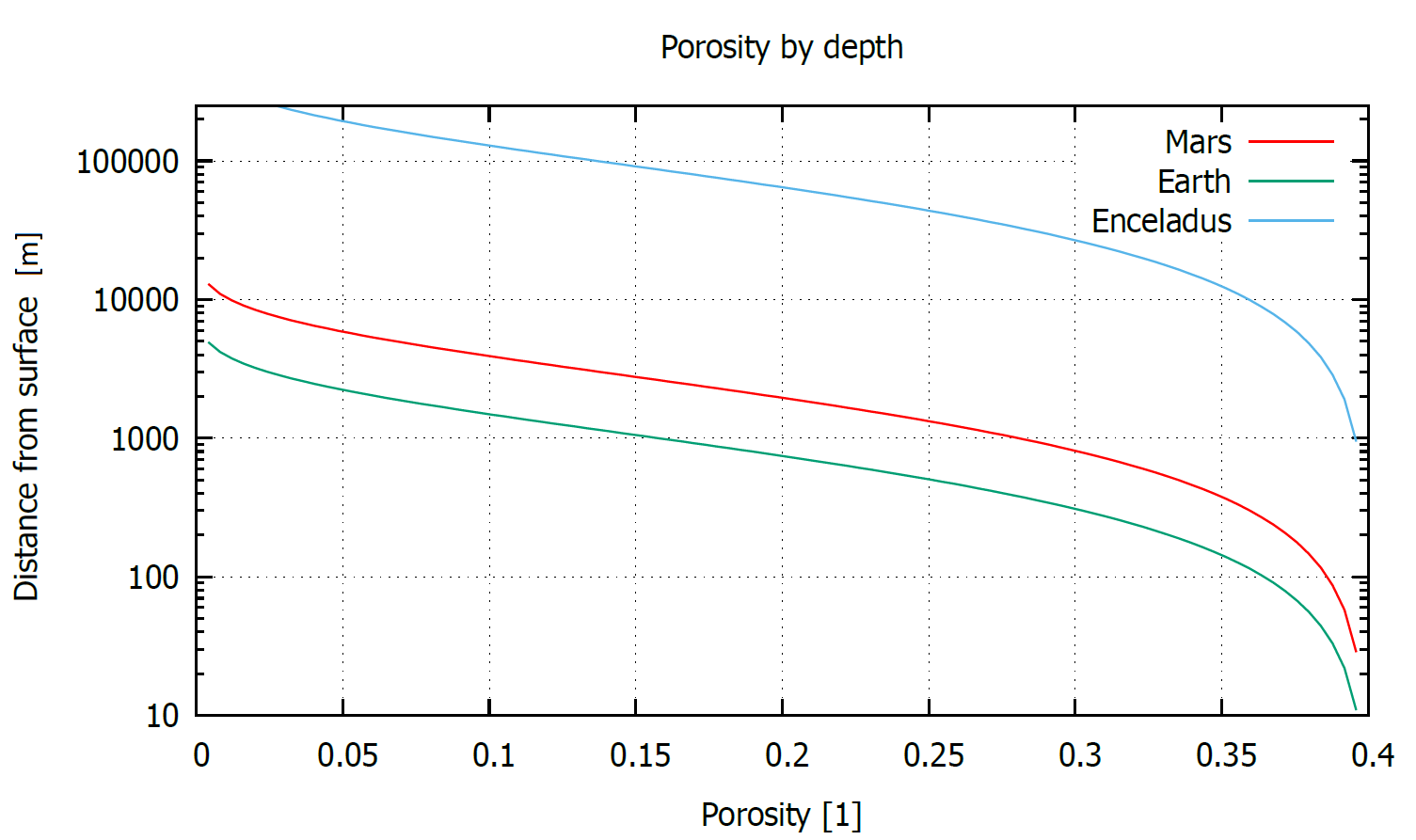}
    \caption{Porosity profiles of Earth, of Mars and of Enceladus at different depths. The curve of Enceladus is based on a surface porosity of 40\%, while the upper boundary of the graph is set to 190 km, which is equal to the radius of the core of Enceladus.}
    \label{fig:logdepth}
\end{figure}

 A potential porosity profile of Enceladus, of the Martian crust and of the crust of Earth are illustrated in Figure \ref{fig:logdepth}. It is based on a surface porosity of 40\% on each body, which value is consistent with the maximum surface porosity of the carbonaceous chondrite meteorite samples measured by \citet{macke2011density}. This initial value provides an upper limit for all the results of our calculations. The curve of Enceladus represents the empirically calculated value of the porosity.

\begin{figure}
	\centering
	    \includegraphics[width=\columnwidth]{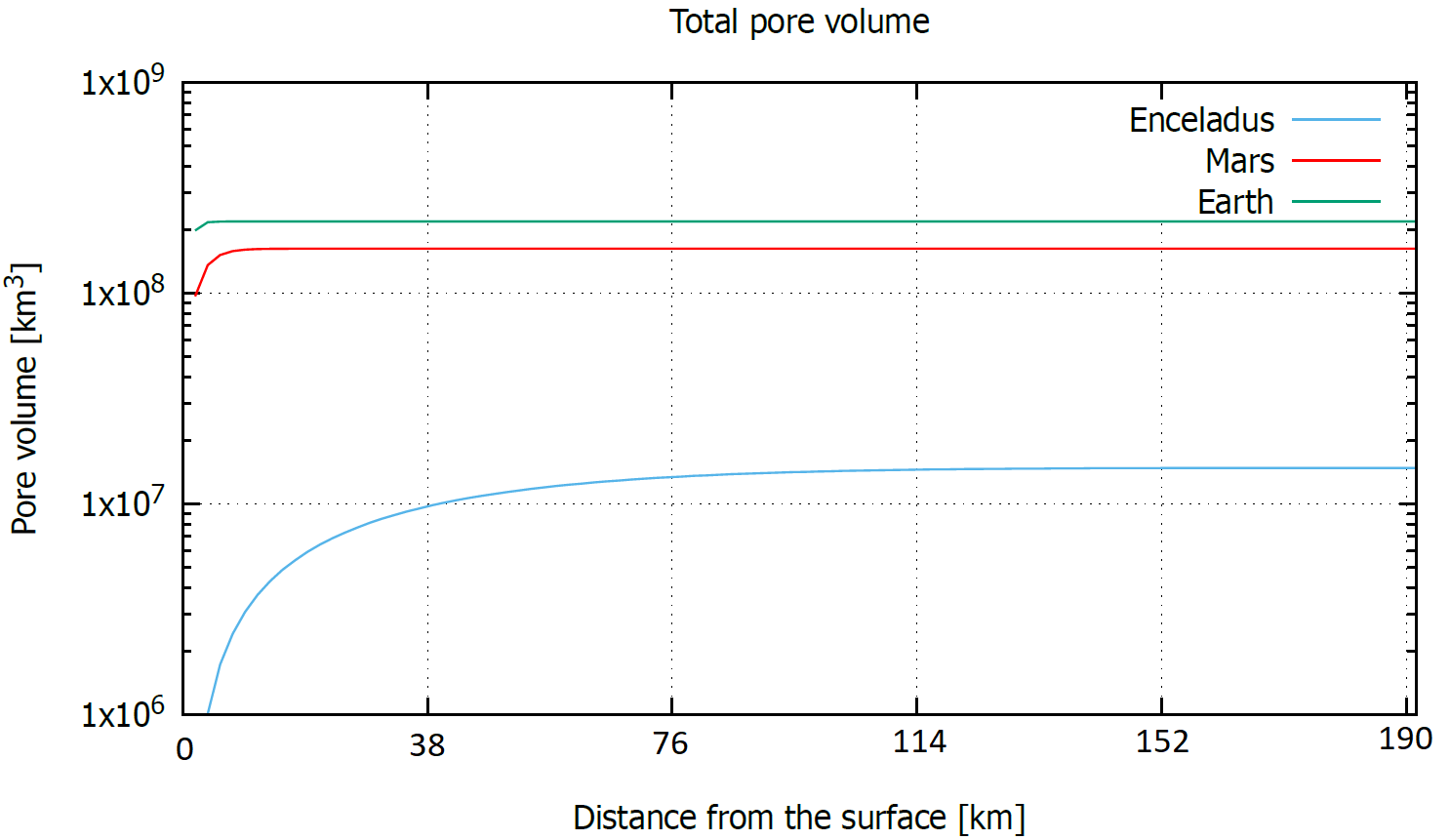}
	    \caption{The total pore volume of each body summarized in a logarithmic graph. As it shows, Earth and Mars compacts very rapidly with depth, therefore they reach their total pore volume value faster than Enceladus. The total pore volume of Enceladus is $1.51*10^7 km^3$, while the same parameter of Earth and Mars is $2.11*10^8 km^3$ and $1.62*10^8 km^3$ respectively.}
	    \label{fig:porevall}
\end{figure}

\begin{figure}
	\centering
	    \includegraphics[width=\columnwidth]{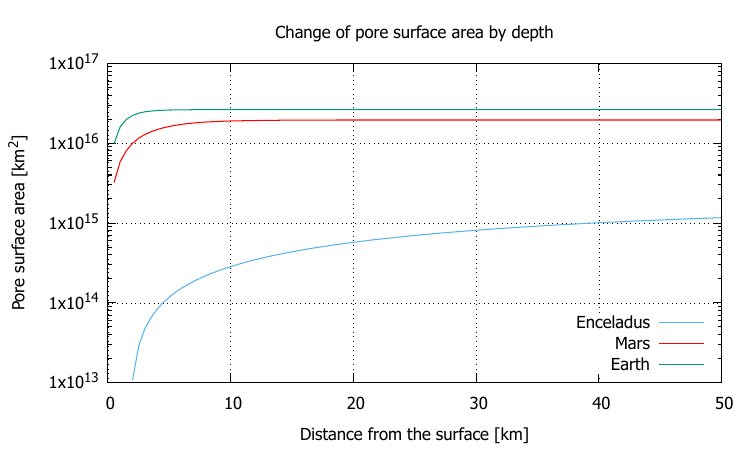}
	    \caption{The total pore surface area of the three investigated planetary bodies summarized in a logarithmic graph. According to our calculations, the pore surface of Enceladus has a maximum value of $1.777*10^{15} km^2$. The same parameter of Mars and Earth has a maximum value of $1.950*10^{16} km^2$ and $2.726*10^{16} km^2$.}
	    \label{fig:poreaall}
\end{figure}

According to our model the self-compaction depth of Earth and Mars is at about 4.5 km and 10.5 km below the surface respectively, meanwhile the interior of Enceladus never compacts fully due to the low mass and the related gravity of the planetary object. The very centre of Enceladus has a porosity of 5\% equal to that of Earth roughly at 2.78 km depth, and Mars roughly at 7.30 km depth. With the results of Equation \ref{eq:totalvolume} and \ref{eq:PoreA} we plotted the total pore volume and total pore surface of Earth, Mars and Enceladus in Figure \ref{fig:V}, \ref{fig:porevall} and \ref{fig:poreaall}.

The results shown on Figure \ref{fig:V} indicate, that there is roughly a 17\% difference in total pore volume between the two methods of porosity calculation. This value is not extremely large, but still considerable, therefore we will not consider the two methods identical in result, but will only perform further calculations using the empirical formula. The comparisons between different planetary bodies and Enceladus are based on the empirical method for each body.

The one order of magnitude difference between Enceladus and the discussed planets is due to both planets having a larger surface than Enceladus. According to our calculations on a global scale the difference in total surface of the bodies outweighs the effect of Enceladus being more porous. Next, we calculated the local concentration of pore volume we used Equation \ref{eq:totalvolume} and plotted the local change in pore volume by depth. The results are presented in the following figure.

\begin{figure}
    \centering
	    \includegraphics[width=\columnwidth]{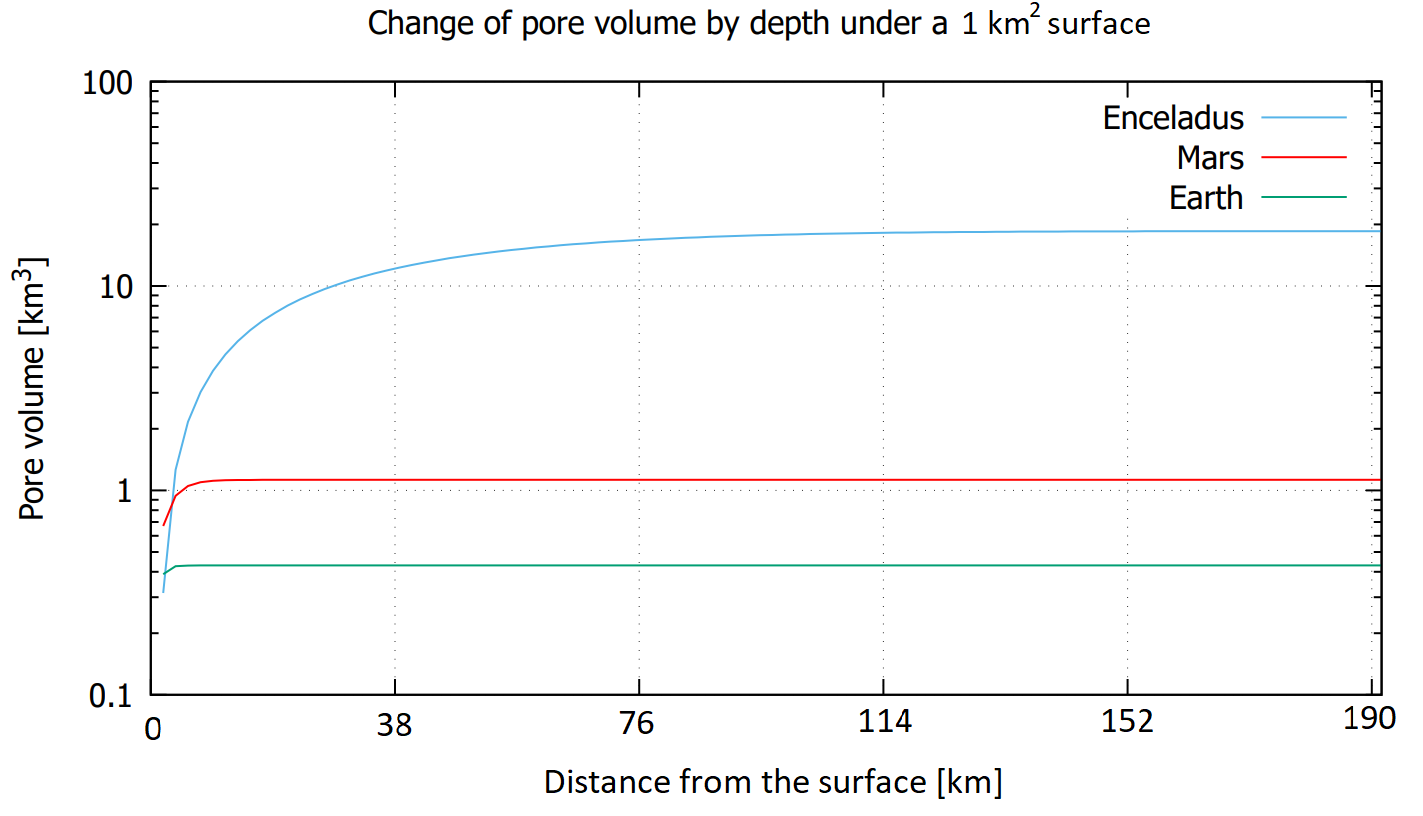}
	    \caption{The graph shows the amount of pore volume present under a unit area of surface on each planetary body.}
    \label{fig:porev1km}
\end{figure}

As shown on Figure \ref{fig:porev1km}, there is about 19.58 $km^3$ pore volume under a 1 $km^2$ area on Enceladus, 1.08 $km^3$ on Mars and only 0.40 $km^3$ on Earth. The results of Figure \ref{fig:porev1km} can be explained by Enceladus having a smaller surface gravity, than Earth and therefore being more porous at every depth except the surface. The difference in the vertical pore distribution might cause and allow a wider storage of liquids, what the localized (like near surface) chemical reactions could utilize during a longer period - however this aspect is complex and requires further evaluation. Using equation \ref{eq:1kmA} we plotted the change of pore surface area by depth under a $1 km^2$ surface for Earth, for Mars and for Enceladus. The results are shown below on a logarithmic scale.

\begin{figure}
    \centering
	    \includegraphics[width=\columnwidth]{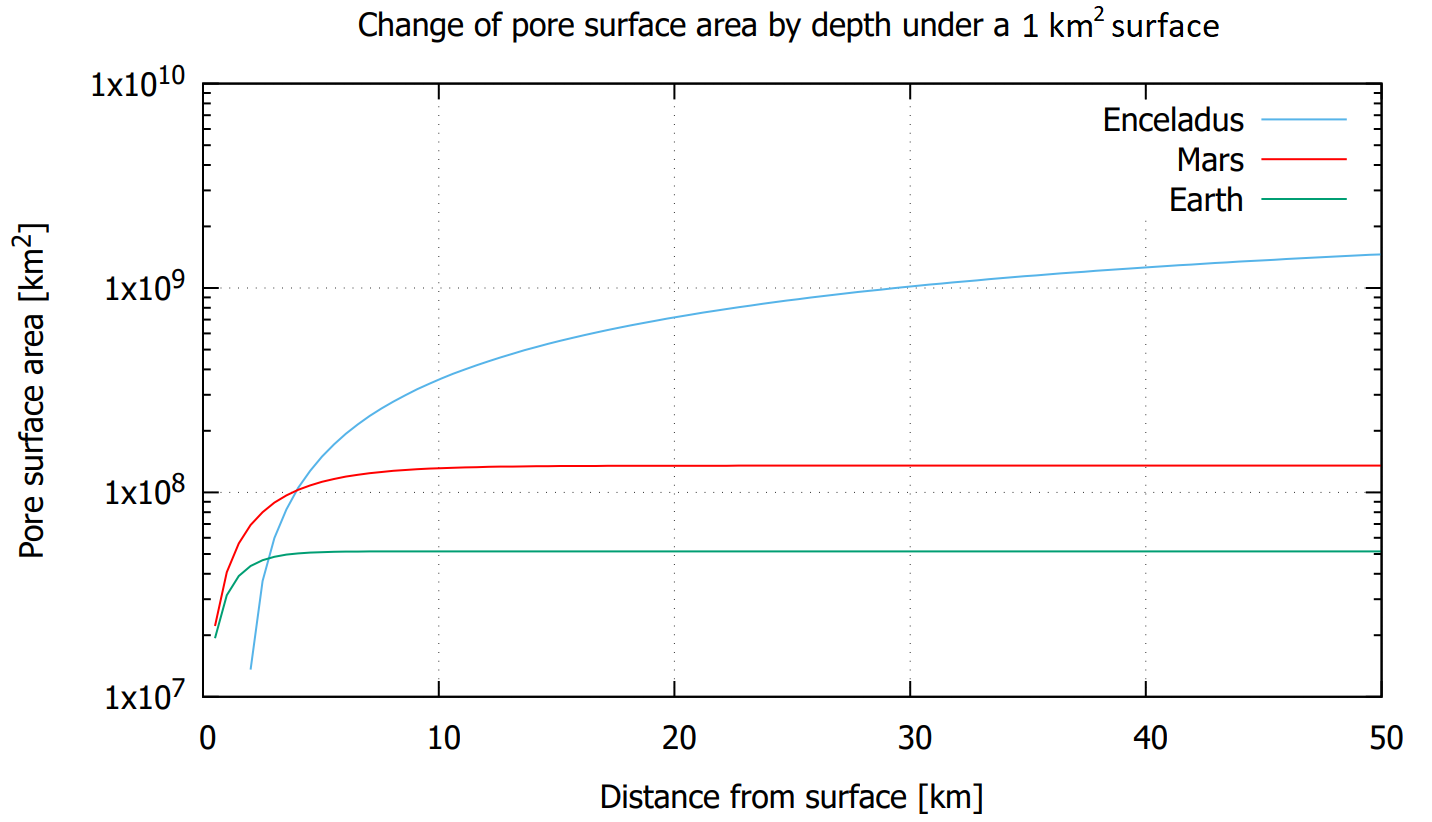}
	    \caption{Enceladus has $1.37*10^9 km^2$ pore surface area under a surface of $1km^2$. Compared to this value, Earth has much less pore surface locally with an estimated $5.07*10^8 km^2$. The maximum value of pore surface area on Mars is calculated to be $1.23*10^8 km^2$.}
    \label{fig:poresurface}
\end{figure}

\begin{figure}
    \centering
	    \includegraphics[width=\columnwidth]{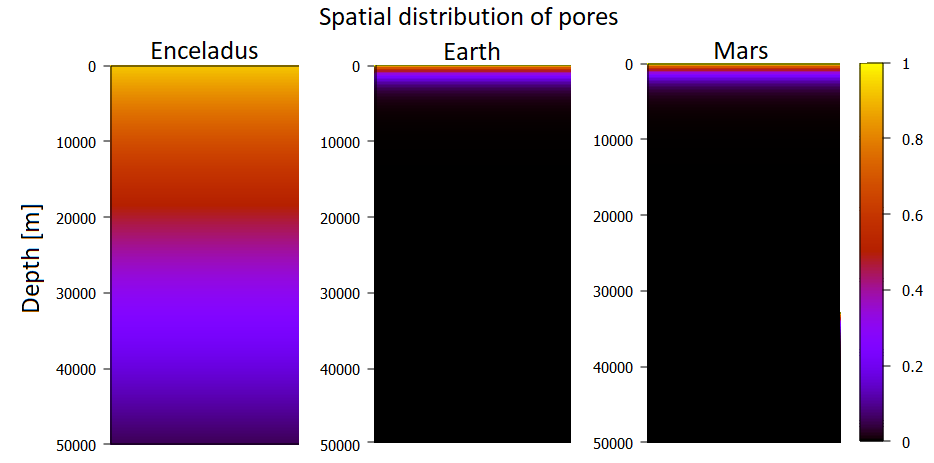}
	    \caption{Spatial concentration map of the pores inside Enceladus (first column), Earth (second column) and Mars (third column)  respectively. For Enceladus we plotted the map from the core to the surface. Advancing downwards, the pore surface area only reaches 50\% of the value of the total pore surface area at the depth of 40 km. We can see that inside Earth and Mars the pores are strongly concentrated near the surface. This is mainly due to the shallow pore compaction depth, which is caused by the high surface gravity of the body. About 50\% of the total pore surface area can be found above the depth of 1000 m.}
    \label{fig:heatmap}
\end{figure}

To better understand the spatial distribution of the pore surface area inside each body, we created maps showcasing the spatial distribution of the total pore surface by depth. These maps are shown in Figure \ref{fig:heatmap}, Note that the vertical axis has the same scale for each body, meaning, that the shown areas of depth stretch far into the interior of Enceladus, while on Earth and Mars it only shows a small section of the crust.

\section{Discussion}
\label{sec:discussion}

The results of our calculations imply that there is a significant numerical difference between the porosity values calculated using the method of experimental equations and the method of interpolating measurement data. Therefore we conclude that the results of these two different methods can not be considered identical and we chose to only model further porosity related parameters using the likely more accurate experimental method.\\

The values presented in Section \ref{sec:results} show significant differences between Enceladus and the investigated rocky planets - Earth and Mars - considering their vertical porosity distribution. Considering the pore compaction depth, Enceladus stays porous even at the very centre, with an upper limit for minimum porosity of 5\%. For comparison, the crust of Earth and Mars have a porosity of 5\% at the depth of 2.36 km, and 6.03 km respectively. The unconsolidated interior of Enceladus results in a structure that allows the hydrothermal circulation of liquid water.
The substantial difference in the pore and pore surface distribution inside he three analysed bodies might have consequences on  the spatial distribution and intensity of chemical reactions. While in the case of Earth and Mars pore volumes and pore surfaces support chemical reactions and related mineral changes only in the topmost shallow subsurface region, however inside Enceladus a sparser distribution is available (both in absolute numerical sense and also regarding the size of the body), which allows the access and contribution to reactions by such elements, which are only available in deeper regions. For example much frequent or abundant events could be reactions with iron, nickel and other heavy nuclei atoms inside pores located anywhere inside Enceladus while for Mars and Earth pores do not exist at depth of the metal segregated core.\\

The loose and permeable inner structure also supports the possibility of a higher amount of tidal heating to be generated through internal friction, therefore it may significantly contribute to the heat budget \citep{2022MNRAS.517.3485K, Neumann_2019}. The results shown in Figure \ref{fig:heatmap} indicate that on Earth and on Mars the pore spaces are highly concentrated near the surface, resulting in no pores below the depth of 5-10 km. On the other hand, on Enceladus pore spaces can be found at every depth of the interior. Therefore the scale of fluid transport and related chemical reactions between water and rock and mineral transport may also be stronger on Enceladus. \\

Compared with recent studies conducted on the inner structure of Enceladus, the models describing the interior of Enceladus as mainly unconsolidated, loose and highly fragmented \citep{choblet2017powering, roberts2015fluffy} are inherently closer to our results, than the ones describing it as mostly or fully compacted \citep{Neumann_2019, prialnik2008growth}. Despite our model being a direct application of experimentally derived general equations on Enceladus, the porosity profile derived by us is similar to the porosity profile derived in the study of \citet{choblet2017powering}, who used the assumed composition and measured density of Enceladus to calculate the porosity. The main similarity between the profiles are the uncompacted interior and the exponential decrease of porosity with depth. The maximum value for the total pore volume and total pore surface area of Enceladus is in agreement with gravity field measurements assuming that the pore spaces are filled with water and a silicate rheology \citep{thomas2016enceladus, vcadek2016enceladus}. Further analysis regarding the effect of these simplifications is certainly welcome. We find the results of \citet{Neumann_2019} and \citet{malamud20161} an underestimation of the porosity related values of Enceladus. The cause of this disagreement could be that the rock type chosen by the authors is less viscous and compacts more rapidly than the actual dominant rock of Enceladus. \\ 

Considering the total pore volume our results are consistent with the existing literature in the case of Earth \citep{chen2020empirical,athy1930density} and in the case of Mars too \citep{clifford2010depth, michalski2018martian, michalski2013groundwater}. With the help of our calculations of the spatial distribution of pores inside the planetary bodies, local processes, such as chemical reactions like serpentinization (mineral conversion of olivine by hydration with heat release), can be better evaluated as their intensity is influenced by the amount of surface on which water and rock can interact \citep{moody1976serpentinization}.\\

There is a wide range of possibilities how this model could be developed further in the future. Beside the application to other moderately small satellites (including Mimas and other objects), methodological development and additional parameters and processes that effect the porosity are planned to be implemented.\\

Internal porosity plays a complex role in the geological evolution and structure of planetary bodies. Here, we only mention some possible implications of our calculations in further studies, connected to porosity dependent geological and geophysical processes inside small to large sized rocky planetary bodies. Although several factors have not been considered here, as a simple early step the gravity driven compaction could already provide useful findings, which should be further improved by additional parameters in the future. Fluid flow is one of the many processes influenced by porosity. With greater porosity values the scale of fluid flow increases, producing a higher amount of solved mineral transport and supporting the homogeneous distribution of locally released internal heat (like localized radioactive elements or differentiation produced heat of the core). The mineral transport inside a body (both in solved or later elsewhere crystallized forms) contributes to a variety of chemical reactions shaping the composition and structure of the interior \citep{czechowski2014some}. Another factor is the pore surface area. With the higher pore surface area, the total reactive surface on minerals enlarges, which can be found through the whole body only in such moderately small sized bodies as Enceladus, where the fractures in the loose interior interconnect the deep interior and the surface regions. The fluid percolation related possibilities might influence the available chemicals for mantle and crust metasomatism \citep{carter2021multi} together with the produced minerals modifying rheological properties \citep{azuma2014rheological}, also partly influencing the redox state of the shallow subsurface regions \citep{lichtenberg2021redox} with many further consequences. Water related reactions might be spatially more restricted from pore fluids in larger bodies -- opposite to smaller bodies where differentiation and pore compaction did not happen, allowing fluid percolation driven increased chemical variability related reactions between almost all type of components and all locations in such bodies.\\

According to studies regarding abiogenesis, there is a high probability, that chemiosmosis played an essential role in the development and evolution of simple organic structures in the oceans of early Earth \citep{bada2007debating}. Chemiosmosis is the movement of ions across a semipermeable membrane bound structure, down their electrochemical gradient \citep{bada2007debating}. Earlier cells might have had a leaky membrane and been powered by a naturally-occurring redox gradient near places of high hydrothermal activity \citep{holm2005hydrothermal}. Hydrothermal activity is present on Enceladus too, as recent studies on the moon show \citep{waite2017cassini}. This process is influenced by the porosity of the rock at  the ocean floor, as discussed above, thus our calculations support further studies aimed to investigate the geological factors contributing to the habitability potential of planetary bodies at subsurface rock-water interfaces.

\section{Conclusion}
\label{sec:conclusion}

In this work, we present the application of experimentally derived equations of porosity on a medium sized icy moon, Enceladus. We compared these results with the calculated porosity values of Earth and of Mars to get a general overview of the effect of the varying surface gravity and radii on the porosity profile of planetary bodies. With this method we derived an upper limit for the pore compaction depth of 4.5 km and 10.5 km for Earth and for Mars respectively, meanwhile Enceladus does not compact fully even at the centre (at 252 km depth) and preserves a minimum porosity of 5\% there. The total pore volume of Enceladus, of Earth and of Mars are estimated to be $1.51*10^7 km^3$, $2.11*10^8 km^3$ and $1.62*10^8 km^3$ respectively, indicating roughly the same scale of capacity in holding various fluids in their interior.\\

We also calculated the values of pore volume under a given 1 $km^2$ area of surface for each body. The result for Enceladus is 18.62 $km^3$, 1.05 $km^3$ for Mars and 0.41 $km^3$ for Earth - indicating a vertically more distributed fluid volume down to the core for Enceladus. Enceladus has $2.89*10^9 km^2$ of pore surface area under a given surface of $1 km^2$, while Earth has about $5.02*10^7 km^2$ and Mars has $1.26*10^8 km^2$ according to our model. The results of our calculations are in agreement with structural evolution models that use a different variety of initial parameters for the calculation of porosity resulting in a loose, sandpile-like inner structure \citep{choblet2017powering, roberts2015fluffy}.\\

The porous interior obtained here contributes to the understanding of Enceladus, as porosity is relevant for further processes both in the interior and at the surface. For example lare scale hydrothermalism in a porous core can contribute to the proposed global sale asymmetric ice thickness \citep{travis2015keeping, roberts2015fluffy}. Porosity also influences the geochemical processes inside Enceladus, which effect can be investigated too in future studies. Finally, as this model provides a coherent porosity structure for Enceladus and other rocky planetary bodies, it can also be used to estimate the effect of porosity on the habitability potential through influencing the scale of certain chemical processes. The coherence between the results of the theoretical evolution models in the literature \citep{choblet2017powering, roberts2015fluffy} and the experimental equations will also promote the use of these methods in the exploration of the porosity structure of planetary bodies in general.

\section*{Acknowledgements}

Imre Kisvárdai thanks the financial support provided by the undergraduate research assistant program of Konkoly Observatory. The authors also acknowledge the helpful suggestions from the anonymous reviewer.

\section*{Data Availability}

The data acquired and analysed for this study will be shared on reasonable request to the corresponding author.



\bibliographystyle{mnras}
\bibliography{mnras2} 




\appendix


\bsp	
\label{lastpage}
\end{document}